\documentclass{elsart}

\usepackage{epsfig}

\usepackage{amssymb}

\def\LA{\left\langle}
\def\RA{\right\rangle}
\def\LP{\left(}
\def\RP{\right)}

\begin{document}

\begin{frontmatter}



\title{Gibbs versus non-Gibbs distributions in money dynamics}


\author[label1]{Marco Patriarca},
\ead{marco@lce.hut.fi}
\author[label2]{Anirban Chakraborti}, and
\ead{anirban@bnl.gov}
\author[label1]{Kimmo Kaski}
\ead{kimmo.kaski@hut.fi}

\address[label1]{Complex Systems Group,
Laboratory of Computational Engineering,\\
Helsinki University of Technology,
P.O.Box 9203, 02015 HUT, Finland}

\address[label2]{Department of Physics, Brookhaven National 
Laboratory, \\
Upton, New York 11973, USA}

\begin{abstract}
We review a simple model of closed economy,
where the economic agents make money transactions and a 
saving criterion is 
present.
We observe the Gibbs distribution for zero saving propensity,
and non-Gibbs distributions otherwise.
While the exact solution in the case of zero saving 
propensity is already known 
to be given by the Gibbs distribution,
here we provide the explicit analytical form of the 
equilibrium distribution
for the general case of nonzero saving propensity.
We verify it through comparison with numerical data
and show that it can be cast in the form of a Poisson 
distribution.

\end{abstract}

\begin{keyword}
Econophysics \sep money dynamics \sep Poisson distribution \sep Gibbs 
distribution
\PACS 89.65.Gh \sep 87.23.Ge \sep 02.50.-r
\end{keyword}
\end{frontmatter}

\section{Introduction}

It is known that the higher end of the distribution of income $f(m)$
follows the Pareto law \cite{Pareto:1897}, 
$f(m) \propto m^{-1-\alpha}$,
where $m$ is the income (money) and the exponent $\alpha$ has a value
in the interval $1$ and $2$  
\cite{Levy:97,Dragulescu:2001a,Reed:2002,Aoyama:2003}.
An explanation of the Pareto law, in terms of the laws regulating
the system micro-dynamics,
should take into account its basic constituents, i.\,e. the trading agents,
as well as the criteria used to carry out the economic transactions. Several 
studies have been made to provide an explanation (see Ref. \cite{Slanina:2003} for a brief summary and more references). In this respect, it is of general interest to study some simple systems of closed economy,
which can be either solved exactly or simulated numerically,
in order to investigate the relation between the micro-dynamics
and the resulting macroscopic money distribution \cite{Chakraborti:2000a,Dragulescu:2000a,Chakraborti:2002a,Hayes:2002a,Chatterjee:2003a,Das:2003}. 
In this paper we consider the generalization of a simple model of money conserving economy,
realized by introducing a criterion of saving in the transaction law,
through the saving propensity $\lambda$.
We study numerically its asymptotic money distribution as a function of the model parameters.
We show that it is not a Gibbs distribution and,
by direct comparison with numerical data, 
that the corresponding analytical solution has the form of a Poisson distribution.

\section{Model}
\label{model}

In the simple model considered \cite{Dragulescu:2000a},
$N$ agents can exchange money in pairs between themselves.
For the sake of simplicity we assume that  
all the agents are initially assigned the same money amount $m_0$,
despite this condition is not restrictive for the following results.
Agents are then let to interact.
At every ``time step'', a pair $(i,j)$ is randomly chosen
and the transaction takes place.
During the transaction, the agent money amounts $m_i$ and $ m_j$ 
undergo a variation, $m_i \to m_i'$ and $m_j \to m_j'$.
Money is assumed to be conserved during the transaction, so that
\begin{equation}
  m_i + m_j = m_i' + m_j' \ .
  \label{conservation}
\end{equation}
In this basic model, $m_i'$ and $m_j'$ are obtained 
through a random reassignment of the total money $(m_i + m_j)$,
\begin{eqnarray}
  m_i' &=& \epsilon \, (m_i + m_j) \ ,
  \nonumber \\
  m_j' &=& (1-\epsilon) (m_i + m_j) \ ,
  \label{basic}
\end{eqnarray}
where $\epsilon$ is a random number, 
extracted from a uniform distribution in the interval $(0,1)$.
Notice that this model of dynamics, as well as its variations considered in the following, 
ensures that agents have no debts after the transaction, 
i.\,e. they are always left with a money amount $m \ge 0$.
It can be shown that, merely as a consequence of the conservation law (\ref{conservation}),
the system relaxes toward an equilibrium state characterized by 
a Gibbs distribution~\cite{Chakraborti:2000a,Dragulescu:2000a,Chakraborti:2002a},
\begin{equation}
  f(m) = \beta \exp \LP - \beta m\RP \ ,
  \label{Gibbs}
\end{equation}
where $\beta = 1/\LA m \RA$ represents the inverse average money and
$\LA m \RA=\sum_i m_i/N \equiv m_0$.
This means that after relaxation, the majority of the agents has a very small amount of money,
while the number of richest agents -- e.\,g. those with $m$ larger than a given value $m'$, 
as well as the fraction of the total money they own,
exponentially decreases with $m'$.
The Gibbs distribution (\ref{Gibbs}) has been shown to represent a robust equilibrium state,
reached independently of the initial conditions also 
in generalized models, such as those involving multi-agent transactions.

However, if a saving criterion is introduced \cite{Chakraborti:2000a,Chakraborti:2002a},
i.e. agents save a fraction $\lambda$ -- the saving propensity --
of the money they have before the transaction is made,
the shape of the equilibrium distribution changes dramatically.
The conservation equation (\ref{conservation}) still holds,
but the money to be shared in a transaction between
the $i$-th and the $j$-th agent is now $(1-\lambda)(m_i+m_j)$.
Then Eqs.~(\ref{basic}) are thus modified,
\begin{eqnarray}
  m_i' &=& \lambda m_i + \epsilon (1-\lambda) (m_i + m_j) \ ,
  \nonumber \\
  m_j' &=& \lambda m_j + (1-\epsilon) (1-\lambda) (m_i + m_j) \ .
  \label{sp1}
\end{eqnarray}
These equations can also be rewritten in the following way,
\begin{eqnarray}
  m_i' &=& m_i + \Delta m \ ,
  \nonumber \\
  m_j' &=& m_j -  \Delta m \ ,
  \nonumber \\
  \Delta m &=& (1-\lambda) [\epsilon m_j - (1-\epsilon) m_i] \ ,
  \label{sp2}
\end{eqnarray}
which clearly shows how money is conserved during the transaction.

\begin{figure}[ht]
\centering
\includegraphics[width=3.2in]{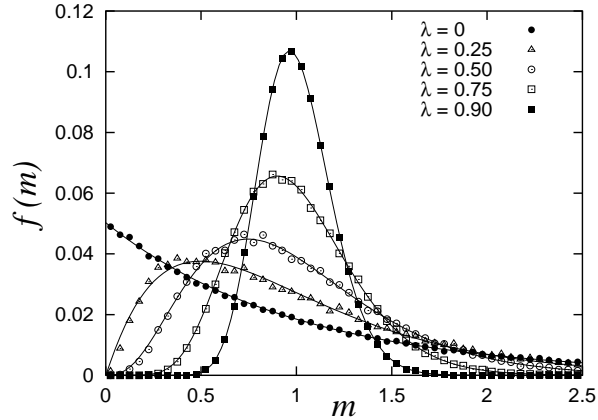}
\caption{
Equilibrium money distributions for different values of the saving propensity $\lambda$,
in the closed economy model defined by Eqs.~(\ref{sp2}).
The continuous curves are the fitting functions, defined in Eq.~(\ref{NGibbs}).
}
\label{fig1}
\end{figure}

We performed numerical simulations, for various values of $\lambda$,
of a system with $N=500$ agents.
In each simulation a sufficient number of transactions,
as far as $10^7$, depending on the value of $\lambda$,
was used in order to reach equilibrium.
The final equilibrium distributions for a given $\lambda$,
obtained by averaging over $1000$ different runs,
are shown in Fig.~\ref{fig1}.
\begin{figure}[ht]
\centering
\includegraphics[width=3.2in]{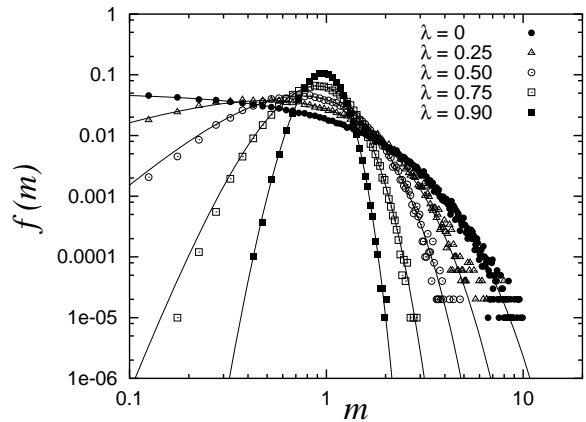}
\caption{
As in Fig.~(\ref{fig1}), but on a double logarithmic scale.
}
\label{fig2}
\end{figure}

\section{Fitting}
\label{fitting}

The exact solution for the case $\lambda=0$ is known to be given 
by the Gibbs distribution, Eq.~(\ref{Gibbs}).
Here we give the corresponding exact solution for an generic value of $\lambda$,
with $0 < \lambda < 1$.
This solution was found by fitting the results of numerical simulations
and it turns out to fit extremely well all data.

It is convenient to introduce the reduced variable
\begin{equation}
  x = \frac{m}{\LA m \RA} \ ,
  \label{x}
\end{equation}
the agent money in units of the average money $\LA m \RA$,
and the parameter
\begin{equation}
  n(\lambda) = 1 + \frac{3 \lambda}{1 - \lambda} \ .
  \label{n}
\end{equation}
We found that the money distributions, for arbitrary values of $\lambda$,
are well fitted by the function
\begin{equation}
  P(x) = a_n x^{n-1} \exp\LP - n x \RP\ ,
  \label{NGibbs}
\end{equation}
where $x$ and $n$ are defined in Eqs.~(\ref{x}) and 
(\ref{n}), respectively\footnote{An excellent fitting is
also obtained if the variable in the exponential is raised to
a power $c$, $\exp(-nx) \to \exp(-nx^c)$, where $c$ is an 
additional parameter and the value of $c$ is close to one 
for all values of $\lambda$. Here we assume $c \equiv 1$, 
since the corresponding fitting is good.}. Using
normalization conditions, the prefactor is easily shown to 
be given by
\begin{equation} 
  a_n = \frac{n^{n}}{\Gamma(n)} \ ,
  \label{a}
\end{equation}
where $\Gamma(n)$ is the Gamma function.

\begin{figure}[ht]
\centering
\includegraphics[width=2.5in]{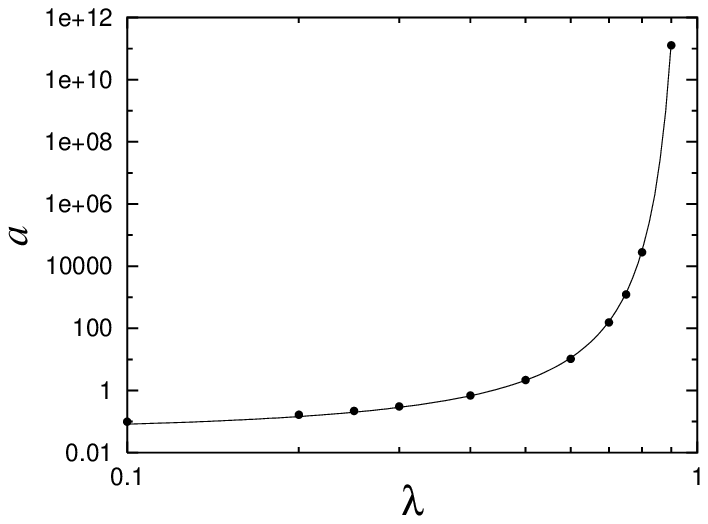},
\includegraphics[width=2.5in]{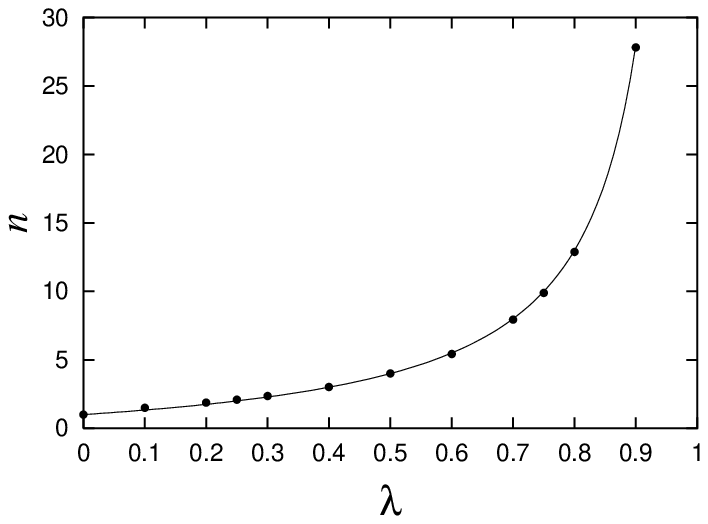}
\caption{
The parameters $a_n$ (left) and $n$ (right) versus $\lambda$
obtained from numerical data (dots) and the corresponding
analytical formulas (continuous curves) 
given by Eqs.~(\ref{a}) and (\ref{n}), respectively.
}
\label{figpar}
\end{figure}

The fitting curves for the distribution (continuous lines) 
are compared with the numerical data in Fig.~\ref{fig1}.
The fitting describes the distribution also at large values 
of $x$,
as shown by the logarithmic plots in Fig.~\ref{fig2}.
The numerical values of the parameters $a_n$ and $n$ are 
compared with the respective
fitting functions (\ref{a}) and (\ref{n}) in 
Fig.~\ref{figpar}.
The distribution function (\ref{NGibbs}) still contains an 
exponential factor
$\exp(-nx)$, similar to that of the Gibbs distribution, but 
the average value
is now rescaled by $n$. 
The power $x^{n-1}$ qualitatively changes the Gibbs 
distribution into 
a curve with a maximum at $x>0$, i.\,e. with a mode 
different from zero.

It is to be noticed that by introducing the rescaled variable
\begin{equation} 
  x_n = nx \equiv \frac{m}{\LA m \RA/n} \ ,
  \label{xn}
\end{equation}
and the corresponding probability density $P_n(x_n) = dF(x)/dx_n \equiv P(x)/n$,
where $F(x)$ is the cumulative function,
and using the explicit expression of $a_n$, Eq.~(\ref{a}),
the distribution (\ref{NGibbs}) becomes
\begin{equation}
  P_n(x_n) = \frac{1}{\Gamma(n-1)} \, x_n^{n-1} \exp \LP - x_n \RP \ ,
  \label{Poisson}
\end{equation}
which reduces to the Poisson distribution for integer values of $n$.

\section{Conclusions and discussion}
\label{conclusions}

We have studied a generalization of the simple closed economy model,
in which a random reassignment of the agent money takes place,
by introducing a saving propensity $\lambda > 0$.
We have empirically obtained the corresponding exact analytical solution
from a fitting of the numerical data.
The distribution naturally lends itself to be interpreted
as a Poisson distribution $P(n,x_n)$ for the reduced variable $x_n = m/(\LA m \RA/n)$.
The parameter $n = 1 + 3\lambda/(1-\lambda)$ is in principle continuous
but it can vary between $1$ and $\infty$ when $\lambda$ varies between $0$ and $1$. This result raises the problem of a 
more rigorous derivation of the solution
as well as of a deeper physical interpretation of the result.

\ack
This work was partially supported by the Academy of Finland,
Research Center for Computational Science and Engineering,
project no. 44897 (Finnish Centre for Excellence Program 2000-2005).
The work at Brookhaven National Laboratory was carried out under Contract No.
DE-AC02-98CH10886, Division of Material Science, U. S. Department of Energy.






\begin{thebibliography}{08}
\bibitem{Pareto:1897}V. Pareto, {\it Cours d'economie politique}, Lausanne and Paris, 1897.
\bibitem{Levy:97}M. Levy, S. Solomon, Physica A 242 (1997) 90.
\bibitem{Dragulescu:2001a}A. Dr\u agulescu, V. M. Yakovenko, Physica A 299 (2001) 213.
\bibitem{Reed:2002}W. J. Reed, B. D. Hughes, Phys. Rev. E 66 (2002) 067103.
\bibitem{Aoyama:2003}H. Aoyama, W. Souma, Y. Fujiwara, Physica A 324 (2003) 352.
\bibitem{Slanina:2003}F. Slanina, arXiv:cond-mat/0311235 (2003).
\bibitem{Chakraborti:2000a}A. Chakraborti, B. K. Chakrabarti, Eur. Phys. J. B.
17 (2000) 167.
\bibitem{Dragulescu:2000a}A. Dragulescu and V. M. Yakovenko, Eur. Phys. J. B 17
(2000) 723.
\bibitem{Chakraborti:2002a}A. Chakraborti, Int. J. Mod. Phys. C 13 (2002) 1315.
\bibitem{Hayes:2002a}B. Hayes, American Scientist 90 (2002) 400.
\bibitem{Chatterjee:2003a}A. Chatterjee, B. K. Chakrabarti, S. S. Manna, arXiv:cond-mat/0301289 (2003).
\bibitem{Das:2003}A. Das, S. Yarlagadda, arXiv:cond-mat/0310343 (2003).
\end{thebibliography}
\end{document}